\documentclass[b5paper,twoside]{jpconf}  %Default declaration%
\usepackage{graphicx}

\begin{document}

\title[The KVN+VERA Joint Array]{Early Phase Observations of the KVN+VERA Joint Array}  

\author[Sawada-Satoh et al. ]{S. Sawada-Satoh$^1$, VERA team$^1$ and KVN team$^2$}

\address{$^1$Mizusawa VLBI Observatory, NAOJ, Oshu,  Iwate 023-0861, Japan}
\address{$^2$Korea Astronomy and Space Science Institute, Daejeon 305-348, Korea}
%\address{$^3$Space Physics and Energetic Particle Lab, Department of Physics, Faculty of Science, Mahidol University, Bangkok 10400, Thailand}
%\address{$^4$Thailand Center of Excellence in Physics, CHE, Ministry of Education, 50200, Thailand}

\ead{satoko.ss@nao.ac.jp} 

\begin{abstract}
The KVN+VERA array is a joint VLBI project of seven VLBI stations 
spread throughout Korea and Japan. 
Since the first fringe detection in 2008, the early phase observations 
of the KVN+VERA have been carried out every several months.
Currently, two observing bands of 22 and 43 GHz are available. 
We are aiming for early realization of science observations 
with the 1-Gbps recording system from 2012. 
\end{abstract}

\section{The KVN+VERA array}

VERA (VLBI Exploration of Radio Astrometry), 
led by NAOJ in cooperation with several Japanese universities,  
is a VLBI array to aim for obtaining 3-dimensional map of the Milky Way galaxy.  
It consists of four 20-m antennas located 
at Mizusawa, Iriki, Ogasawara and Ishigaki in Japan, 
to achieve baselines longer than 1000 km up to 2300 km. 

KVN (Korean VLBI Network), promoted by KASI,  
is the first dedicated mm-wavelength VLBI array, 
which consists of three 
21-m antennas located at Yonsei (Seoul), Ulsan and Tamna (Jeju island). 
The baseline length ranges between 300 and 500 km.

On the basis of the VLBI collaboration agreement between KASI and NAOJ, 
we have started the joint observation with the KVN+VERA array. 
The KVN+VERA array complements baseline length range 
up to 2300 km fully, and can achieve a good imaging quality. 
The early phase observations have been carried out using the K4/VSOP 
terminal at 128 Mbps sampling, alternatively. 
The experiments history for the early phase test observations 
with the KVN+VERA array is listed in Table~\ref{his}.

\begin{table}
\caption{\label{his} The experiments history}
\begin{center}
\begin{tabular}{llll}
\br
Date & Bands & Antennas & Notes \\
\mr
2008 nov & 22 & KVN-Yonsei, VERA & The fringe detection test. \\
2009 mar & 43 & KVN-Yonsei, VERA & The fringe detection test. \\
2009 oct & 22, 43 & All of the KVN+VERA & The fringe detection test. \\
2010 apr & 22, 43 & All of the KVN+VERA & The first imaging test. \\
\br
\end{tabular}
\end{center}
\end{table}

\begin{figure}
\begin{center} 
\includegraphics[width=75mm]{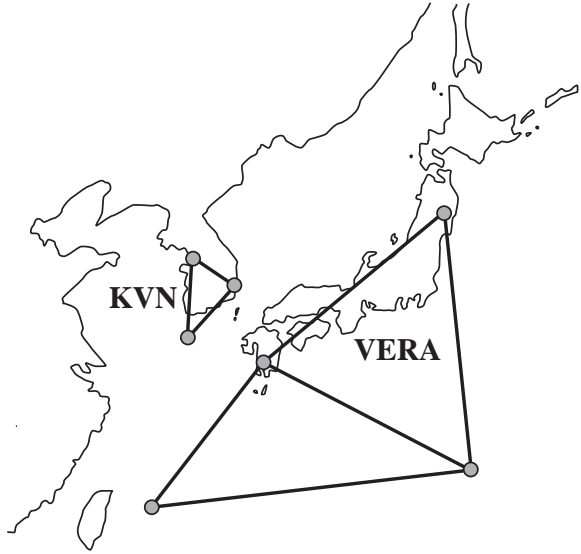} 
\end{center}
\caption{\label{label}
The location of the KVN and the VERA antennas. 
} 
\end{figure}

\section{Imaging ability}

\subsection{Continuum source : 1928+738}

\begin{figure}
\begin{center} 
\includegraphics[width=120mm]{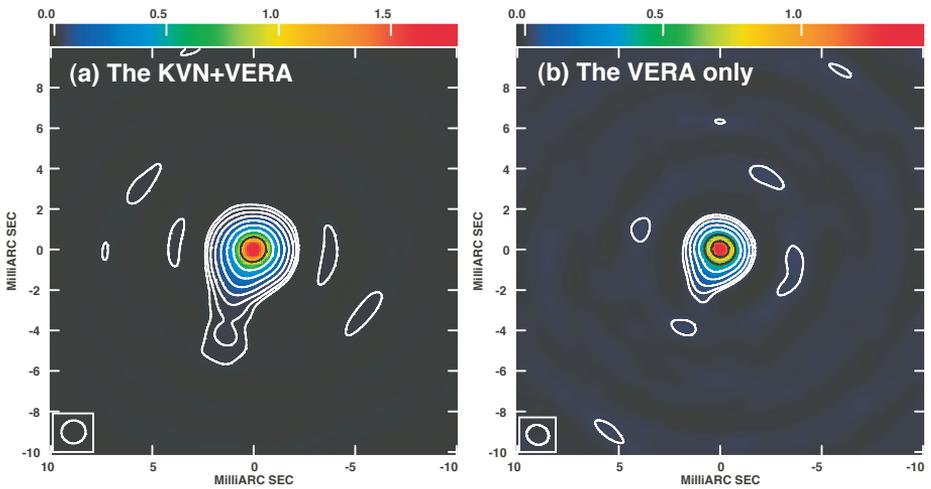} 
\end{center}
\caption{\label{1928}
Two VLBI images of 1928+738 obtained by (a) the KVN+VERA and 
(b) the VERA only.  Achieved image rms level is (a) 2.67 mJy beam$^{-1}$
and (b) 5.34 mJy beam$^{-1}$, respectively. 
} 
\end{figure}

The imaging test observations at 22 GHz 
toward a continuum radio source 1928+738 were carried out 
on 26 January 2011 with the KVN+VERA array. 
%On-source time was 24 hours because 1928+738's DEC is enough high. 
This source is known to exhibit an extended jet structure along the north-south direction \cite{hum92}.   
The KVN+VERA image succeeds to detect the extended jet structure toward south, 
although the VERA image at 22 GHz does not show the structure clearly. 
(Figure~\ref{1928}).  

\subsection{SiO maser v=2 in Orion KL}

The Orion KL region contains strong SiO maser emissions, which 
have been imaged with VLBI techniques. 
Those images show that 
the 43 GHz SiO masers are distributed in four regions that make X shape 
\cite{gre98}\cite{doe99}\cite{kim08}. 
Distribution and velocity field of SiO $v$=2 maser spots in the image
 obtained by the KVN+VERA array are consistent with the image in 2008 with the VERA
 \cite{kim08}. 
The number of detected maser spots ($> 10 \sigma$) are around 250, 
and it is twice more than the VERA image in spite of shorter observation time.

\section{KJJVC and observing modes}

KASI and NAOJ have been developing compatible data acquisition system 
and a common correlator, KJJVC (Korea-Japan Joint VLBI Correlator).
KJJVC is able to process 16 stations, at the maximum sampling rate of 8 Gbps/station. 
KJJVC can accept data from several different VLBI playback systems such as Mark 5b, 
VERA2000 and K-5, and correlate Gbps sampling data between 
the KVN Mark 5b and the VERA2000 backend terminals \cite{oh11}. 
Possible observational modes for the KVN+VERA array are shown 
in Table~\ref{mode}.

\begin{table}
\caption{\label{mode} The observational modes for the KVN+VERA}
\begin{center}
\begin{tabular}{cccccc}
\br
Mode & IF Num. & Bandwidth & Bits & Max data rate &  Compatible  \\
	&	&  (MHz)	&  &(Mbps) & VERA modes	 \\
\mr
1 & 1    & 256 & 2 & 1024 & -- \\
2 & 1,2 & 128 & 2 & 1024 & VERA1 \\
3 & 1,2,3,4 & 64 & 2 & 1024 & VERA2 \\
4 & 1,2,3,4 & 32 & 2 & 1024 & VERA4 \\
5 & 1,2,3,4 & 16 & 2 & 1024 & VERA7, VERA9 \\
  &	&	& 	& 	& GEO1, GEO2 \\
6 & 1,2,3,4 & 8 & 2 & 512 & GEO3, GEO4 \\
7 & 1,2,3 & 64/128 & 2 & 1024 & VERA3 \\
8 & 1,2,3,4 & 32/64/128 & 2 & 1024 & VERA5 \\
9 & 1,2,3,4 & 32/128 & 2 & 1024 & VERA6 \\
10 & 1,2,3,4 & 16/32/128 & 2 & 1024 & VERA8 \\  
\br
\end{tabular}
\end{center}
\end{table}

\section{Current and future works}

The KVN+VERA array adopts a priori amplitude calibration. 
Each KVN and VERA antenna has the chopper wheel of the hot load and 
the system noise temperature is obtained by R-Sky method. 
VERA antennas measure the sky power even during scans, which allow 
frequent system temperature measurements. 
The similar frequent measurement system will be installed to the KVN.

The elevation dependence of the aperture efficiency of the KVN and VERA antennas 
was measured by observing bright H$_2$O and SiO maser sources at 22 and 43GHz.
For both of KVN and VERA, the aperture efficiencies are flat 
at elevation of $> 20^{\circ}$.
The aperture efficiency in low elevation of $< 20^{\circ}$ decreases slightly, 
but this decrease is less than about $10\%$ \cite{vera11}\cite{lee11}.

Currently, several test observations have been scheduled for the KVN+VERA array. 
First, geodetic VLBI observation will be carried out to obtain accurate KVN antenna locations. 
After that, we have several plans for observations in multi-frequency mode. 
The KVN has introduced a simultaneous multifrequency receiver system 
that performs simultaneous observations at four frequencies of 22, 43, 86, and 129 GHz  
\cite{han08}. 
The system allows to calibrate the atmospheric fluctuations 
at 43, 86, or 129 GHz from the visibility phase at 22 GHz \cite{jun11}. 
Korea-Japan joint science WG has started to discuss about the early science observations 
with the KVN+VERA array, 
and the observations will be scheduled end of 2011.  

%acknowledgement
\vspace{5mm}
We are grateful to all members of KVN and VERA for their support 
for the KVN+VERA observations.

\end{document}